\documentclass[aps,floatfix,prc,showpacs,showkeys,amsmath,amssymb,superscriptaddress, twocolumn]{revtex4-1}
\usepackage{graphicx}
\usepackage{xcolor}
\usepackage{ulem}
\usepackage{hyperref}
\usepackage{amssymb}

\usepackage{amsmath,amssymb,graphicx}
\usepackage{graphics}
\usepackage{amssymb}

\renewcommand\sout{\bgroup \color{blue}\ULdepth=-.5ex \ULset}
\begin{document}
\vskip 0.3cm
\title{Light nuclei production in pp and pA collisions in the Baryon Canonical Ensemble}
\author{Natasha Sharma}
\affiliation{Indian Institute of Science Education and Research (IISER) Berhampur, Vigyanpuri, Ganjam, Odisha-760003, India}
\author{Lokesh Kumar}
\affiliation{Department of Physics, Panjab University, Chandigarh 160014, India}
\author{Pok Man Lo}
\affiliation{Institute of Theoretical Physics, University of Wroclaw, Wroclaw PL-50204, Poland}
\author{Krzysztof Redlich}
\affiliation{Institute of Theoretical Physics, University of Wroclaw, Wroclaw PL-50204, Poland}
\affiliation{Polish Academy of Sciences PAN, Podwale 75, PL-50449 Wroc\l{}aw, Poland}
\date{\today}
\begin{abstract}
The increase in  yields of light nuclei  with charged particle
multiplicity, as reported by the ALICE Collaboration at CERN 
in pp and  pPb collisions at the Large Hadron Collider energy  is
investigated in the  thermal hadron resonance gas model.  The model is extended to account for 
exact baryon number conservation. 
The focus is on the production of protons, deuterons,  $^3$He,
and $^3_\Lambda$H.
A very good description of proton and deuteron yields is obtained as a function of 
charged particle multiplicity in the mid-rapidity region using the same 
fixed temperature as in central Pb-Pb collisions. 
The yields of light nuclei
$^3$He and $^3_\Lambda$H, 
though qualitatively explained as a function of multiplicity, are
lower than the model expectation.  
One of the possible reasons could be that for $^{3}$He and $^{3}_{\Lambda}$H, 
the chemical equilibrium is not yet reached at small multiplicities.

\end{abstract}
\maketitle


\section{Introduction}
The production of light nuclei, such as deuteron, triton, $^3$He,
$^3_\Lambda$H, and their antiparticles, at the CERN Large Hadron Collider (LHC) as measured by the ALICE
Collaboration~\cite{ALargeIonColliderExperiment:2021puh,ALICE:2020foi,ALICE:2019bnp,ALICE:2015wav,
ALICE:2017xrp,ALICE:2019fee},  has attracted
considerable attention  recently (for a review see, e.g.,~\cite{D_nigus_2020})   \cite{Andronic:2010qu,Andronic_2018,Cleymans:2011pe,Cai:2019jtk,Donigus:2022haq,Bellini:2018epz,Bellini:2020cbj,Sun:2018mqq,Reichert:2022mek}. 
The thermal model had a spectacular success in predicting their yields~ in central heavy ion collisions with the same thermal parameters as used for all other hadrons  \cite{Andronic:2010qu,Andronic_2018,Cleymans:2011pe}.
Alternate explanations based on the coalescence picture~\cite{Bellini:2018epz,Bellini:2020cbj,Sun:2018mqq,Reichert:2022mek},
contain  in general,  more parameters than the thermal model and a final picture is still not clear.

In this paper, we focus on the quality of the hadron resonance gas (HRG) model description of the yields of light nuclei in small systems  and discuss in detail
their dependence on charged particles multiplicity. The model predictions are compared to
yields that have been measured recently by the ALICE Collaboration in pp and pA collisions in events with  different charge particle multiplicities, ${dN_{ch}/dy}$. 


To this end, we employ the baryon canonical ensemble (BCE) approach for the
reduction (or enhancement) of yields in  the HRG  model~\cite{Vovchenko:2019kes,Cleymans:2020fsc}.
The description of the production of light ions in the thermal model can be 
severely affected by the exact conservation of the baryon number especially
at lower beam energies and small systems.
This important fact was first noted by Hagedorn~\cite{Hagedorn:1971mc} in the context of  anti-$^3$He production in low energy $p\bar{p}$ collisions. There,    the implementation of the exact baryon number conservation in the partition function reduces the yield of anti-$^3$He  by seven orders of magnitude bringing it close to the experimental value.

To take into account the baryon number conservation in the HRG model,  we developed an extension of the THERMUS 
code~\cite{Wheaton:2004qb} to include the baryon canonical ensemble. \
The BCE formulation is especially relevant for describing multi-baryon states produced in events with low values of the accompanying charged particle multiplicity~\cite{Vovchenko:2018fiy}. 

To account for exact baryon conservation in the presence of multi-baryon states, we will follow the procedure outlined in~\cite{Cleymans:2020fsc}
for strange  particles production. We focus on the exact baryon number conservation, where the conservation of all other charges are included in the grand canonical ensemble  with all  chemical potentials  put equal to zero. 
However, the case of  $^3_\Lambda$H yielding the strangeness canonical effects is also included. 
The chemical freeze-out temperature is  fixed
at $T = 156.5$ MeV, a value supported by 
fits to hadronic yields produced in central Pb-Pb collisions at the LHC ~\cite{Andronic_2018,Andronic:2018qqt}, that also coincides with the  chiral crossover temperature  obtained 
by Lattice quantum chromodynamics (LQCD) calculations~\cite{Bazavov:2017dus,Borsanyi:2020fev}.  
Thus, the only parameters left open in the BCE formulation of the HRG model are the volume of the system, $V_A$, and the baryon canonical
volume $V_C$ where exact baryon number conservation $B=0$ is fulfilled.  
We restrict our considerations to pp and pA collisions as the canonical corrections become negligible 
for large systems as seen in Pb-Pb collisions where yields of all  baryons and light nuclei are well described by the HRG model formulated in the  grand canonical ensemble ~\cite{Andronic_2018}. 

\begin{figure}[h!]
\centering
\includegraphics[scale=0.45]{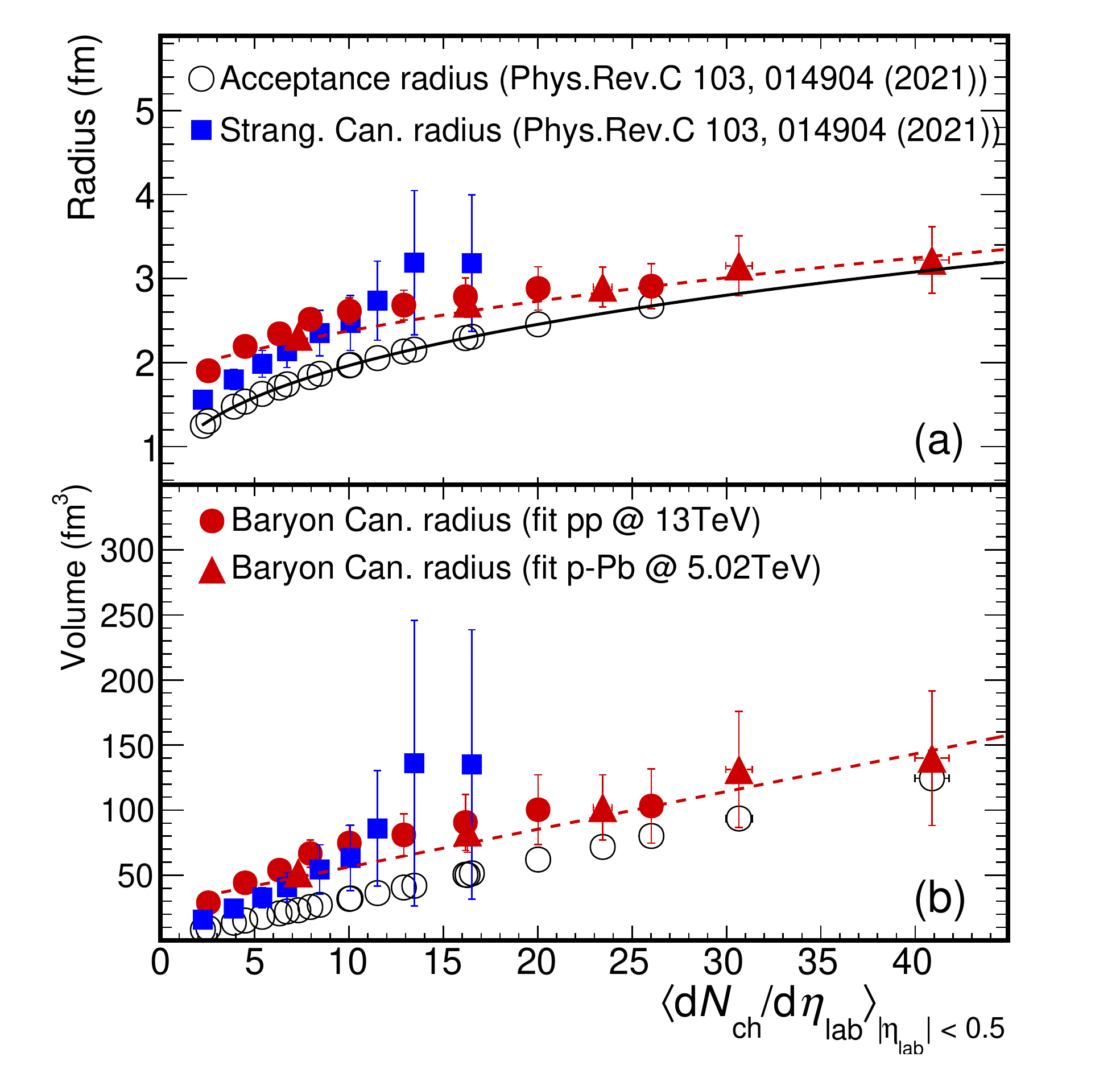}
\caption{Radius (a) and volumes (b) obtained from the thermal model as a function of charged
particle multiplicities.
The solid circles and triangles represent the correlation volume using the BCE formalism. 
The open circles represent the
acceptance volume and the solid squares represent the correlation volume for strange hadrons using the strangeness canonical
ensemble~\cite{Cleymans:2020fsc}. The solid (dotted) line represents the linear fit to the acceptance (baryon canonical correlation) volume as given by Eq.~\ref{eq:volume} (Eq.~\ref{eq:volumec}).}
\label{fig:radius}
\end{figure}

\section{Baryon  canonical ensemble}
The data on light nuclei are obtained in a kinematical region where  net quantum numbers like 
net baryon number and net strangeness are zero as witnessed by the particle-antiparticle symmetry
observed in the central rapidity region  at the LHC. To take this into account we focus on a system which
has zero net baryon number inside a correlation volume $V_C$.
The HRG partition function  is constructed 
by inserting a Kronecker $\delta$ function, thus enforcing zero baryon number:  
\begin{equation}
Z^C_{B=0} = \textrm{Tr}\left[ e^{-H/T}\delta_{(B,0)}\right]  .
\end{equation}
After using the Fourier  representation  of the  delta function,   
the canonical partition function can be written as
\begin{eqnarray}
  Z^C_{B=0}   =  e^{B_0} \frac{1}{2\pi}  
   \int_0^{2\pi}  d\phi 
  \exp \Biggl\{ \sum_{b=1}^{3}  & \Biggl\{  B_be^{ib\phi}   + \Biggr. \Biggr. \nonumber \\
     & \Biggl. \Biggl.  B_{-b}e^{-ib\phi}  \Biggr\}   \Biggr\}
 \label{eq1}
\end{eqnarray}
where
$B_b = \sum_k z_{k,b}$, is the  sum of all single-particle partition
functions, $z_{k,b}~=~V_C n^b_k(T)$, of hadrons with baryon number $b$ while
$Z_{-b}$ is the corresponding  sum for all  antiparticles with
baryon number $-b$. Here, $n^b_k(T)$ is the particle density and $V_C$
is the baryon canonical volume.
The $B_0$ is the partition function containing all hadrons with zero baryon number.
The above expression can be also written as \cite{Braun-Munzinger:2001alt}:
\begin{eqnarray}
  Z^C_{B=0}  = e^{B_0}  \frac{1}{2\pi}\int_0^{2\pi} d\phi 
  \exp \Biggl\{  \sum_{b=1}^{3}  \Biggl\{\sqrt{B_{b}B_{-b}}  ~\times
   \Biggr. \Biggr.
  \nonumber \\
\Biggl. \Biggl.   \Biggl[\sqrt{B_b\over{B_{-b}}}e^{ib\phi}+   \sqrt{B_{-b}\over{B_b}}e^{-ib\phi}\Biggr] \Biggr\} \Biggr\},
\end{eqnarray}
and since at the LHC energies the chemical potentials are all 
equal to zero, one has $B_b=B_{-b}$. 
\par\noindent
Using the well known series expansion for Bessel functions:
\begin{equation}
\exp\left\{\frac{x}{2}\left(t +\frac{1}{t}\right)\right\} = \sum_{m=-\infty}^\infty I_m(x)t^m
\label{eq:canonical3}
\end{equation}
and performing the integral,
one can rewrite the baryon canonical partition function in  Eq.~\eqref{eq:canonical3} as  a series of  
Bessel functions \cite{Braun-Munzinger:2001alt,Braun-Munzinger:2003pwq,Cleymans:1990mn,Hamieh:2000tk}, 
\begin{eqnarray}
Z^C_{B=0} = e^{B_0}\sum_{n,p=-\infty}^{\infty}
a_3^p \, a_2^n \, a_1^{-2n-3p} \,  I_n(x_2) \, \times \nonumber \\ I_p(x_3)   I_{-2n-3p}(x_1). 
\label{eq:partition}
\end{eqnarray}
where
\begin{eqnarray}
a_i &=& \sqrt{B_i/B_{-i}},\\
x_i &=& 2\sqrt{B_iB_{-i}}
\end{eqnarray}
In this way we take into account all baryonic states with
baryon number $\pm 1$, $\pm 2$ and $\pm 3$. Thus, we include deuterons, tritons, $^3$He and their antiparticles but not
particles with higher baryon number like $^4$He and its antiparticle.
Note that in the case considered here, where all chemical potentials are zero, one gets  $a_i=1$ for all $i$.


The resulting yields of particle carrying baryon number $b$  in the 
baryon canonical ensemble  are then given by the following expression:
\begin{eqnarray}
\label{eq:particle}
\langle N_k^b\rangle_A = \,  \frac{z_{k,b}^A }{Z_{B=0}^C}  
\sum_{n,p=-\infty}^{\infty} \, a_3^p \, a_2^n \, a_1^{-2n-3p-b} \,  I_n(x_2) \times \nonumber \\ I_p(x_3) \, I_{-2n-3p-b}(x_1).
\end{eqnarray}
 Furthermore,
 we parameterize $z_{k,b}^A=V_A n_k^b(T)$, where $V_A$ is the volume in the acceptance window.
 We also include the  resonance contributions  to $z_{k,b}$. 
 

It is important to distinguish two volumes in the analysis of yields. 
One {($V_A$)} is   the fireball volume determined by the {experimentally measured charged particle yields within a unit rapidity}, the other one is 
the correlation volume ($V_C$) of exact baryon   conservation;
these two are in general,  not the same \cite{Cleymans:2020fsc,Hagedorn:1971mc,Hamieh:2000tk}. For
  example, in the recent baryon canonical model analysis of net-proton number fluctuations measured  in central Pb-Pb collisions by ALICE Collaboration. it was found, that  the baryon number conservation
is long-range in rapidity and corresponds  to full rapidity coverage, i.e. conservation is global \cite{Braun-Munzinger:2019yxj,Braun-Munzinger:2020jbk}.

From Eq.~\ref{eq:particle}, it is clear  
that the yields  are determined 
by the chemical  freeze-out temperature $T$ and  two volume parameters: 
$V_A$ which appears as an overall factor determining the normalization of the yield and 
$V_C$ of  the  
baryon number conservation which appears in the arguments of the Bessel functions. 


In the following, we apply the above HRG model formulated in the BCE  to describe the  
 yields of protons  and (multi-)baryonic light nuclei and their 
behavior with charged particle multiplicity  as observed by the ALICE Collaboration in different 
colliding systems and collision energies at the LHC.  
  

\section{Results and discussions}


The system under consideration is same as was used to describe  strange particle yields and their dependence on the charged particle multiplicity in pp and pA collisions at the LHC energies \cite{Cleymans:2020fsc,Sharma:2018owb,Sharma:2018jqf}. Thus, we fixed the  freeze-out temperature 
at $T = 156.5$ MeV, and 
the acceptance volume $V_A$ at mid-rapidity  to the values as were  obtained in~\cite{Cleymans:2020fsc}. 
In Fig.~\ref{fig:radius} we  show the radius parameter
and  the corresponding fireball  volume  $V_A$,  and their dependence on $dN_{ch}/d\eta$ from Ref. \cite{Cleymans:2020fsc}. Also shown in this figure is the strangeness canonical volume parameter that was extracted within the HRG model formulated in the canonical ensemble  to successfully describe the strange and multi-strange hadron yields and their observed systematics \cite{Cleymans:2020fsc}.

In the thermal model analysis of the  production yields of  baryons and multi-baryon states in pp and pA collisions one needs to account for the exact baryon number conservation introduced in the BCE in Eq.~\ref{eq:particle}. With the fireball volume in the acceptance window $V_A$ introduced in Fig.~\ref{fig:radius} and the freeze-out temperature from Refs.~\cite{Andronic:2017pug,Andronic:2018qqt} we  are  left with the  baryon canonical volume parameter $V_C$  to fully quantify  data.  
We calculated $V_C$ by fitting pions, protons
and deuterons yields as measured by the ALICE Collaboration for different multiplicity classes in pp and pA collisions  ~\cite{ALICE:2020foi,ALICE:2019bnp,ALICE:2020nkc,ALICE:2016fzo}. 
The results are summarized in Fig.1  where
the solid circles and solid triangles represent the results for pp collisions at $\sqrt{s}=13$ TeV and pPb collisions at
$\sqrt{s_{NN}}=5.02$ TeV, respectively. 


\begin{figure}[h!]
\centering
 \includegraphics[scale=0.38]{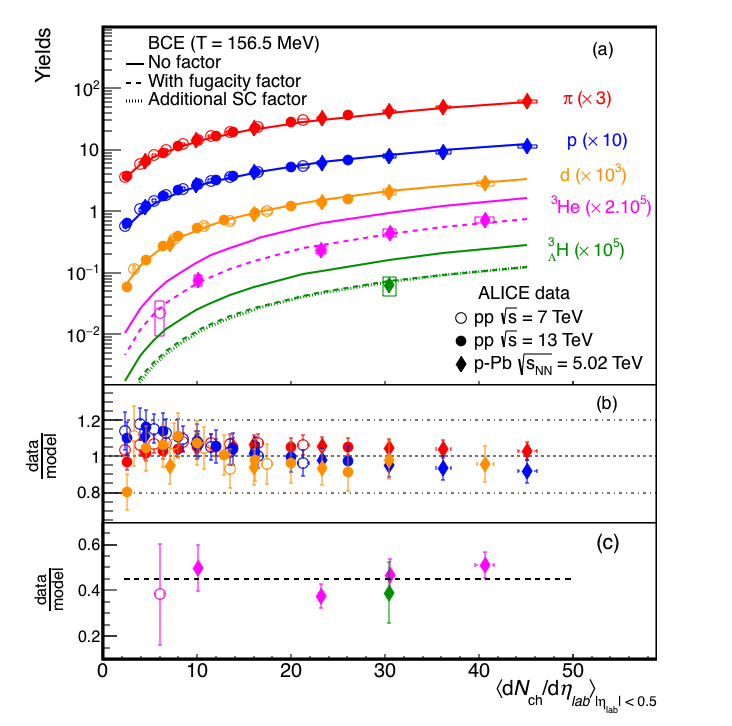}   
\caption{(a) Yields of pion, proton and  light nuclei as a function of the charged particles multiplicities at mid-rapidity~\cite{ALargeIonColliderExperiment:2021puh,ALICE:2020foi,ALICE:2019bnp,ALICE:2015wav,
ALICE:2017xrp,ALICE:2019fee}.
The solid lines represent yields of particles predicted by the BCE
thermal model and the dashed lines are the same predictions scaled by a factor $\gamma$=0.45 obtained from the linear fit in panel. The dotted line represents the effect of exact strangeness conservation in addition to baryon-canonical effects and fugacity factor on $^3_\Lambda$H.
(b) Ratios of data over BCE model for pions, protons, and deuterons. (c) Similar ratios for $^3$He and $^3_\Lambda$H. The dotted line in panel (c) represents the linear fit.
}
\label{fig:yields}
\end{figure}


It is observed, that for low multiplicity, the canonical
volume is only slightly larger than the acceptance volume at
mid-rapidity, indicating that in pp and pPb collisions, the baryon
number conservation is not necessarily extended to the full rapidity
range. This is in contrast to the results of baryon canonical model analysis of net-proton fluctuation data in
central Pb-Pb collisions at the LHC \cite{Cleymans:2020fsc,Braun-Munzinger:2020jbk}.
One of the possible reasons could be we consider here a fully
integrated $p_t$ yields data, while the net-proton fluctuations in central Pb-Pb collisions were
measured over a limited $p_t$ window. Furthermore, the full phase space rapidity distributions of
protons and baryons in pp, pPb, and Pb-Pb collisions are very different and at present not known for events with different $dN_{ch}/dy$.
With future measurements of proton fluctuations 
in pp and pPb collisions, one could verify whether the statistical model in canonical ensemble can provide a  consistent description of particle yields and fluctuations observable. 


The volume deduced from the BCE tends to be also larger than the 
strangeness canonical volume deduced from yields of strange
particles~\cite{Cleymans:2020fsc}.
We note, however, that canonical volume parameters shown in Fig.~\ref{fig:radius} were extracted under the assumption that strangeness and baryon number are conserved independently. 
When the exact baryon and strangeness conservation are  included simultaneously in  the canonical ensemble then yields of  particles    are  not calculated anymore    following  Eq.~\eqref{eq1}. They   are rather obtained from the partition function  described by the double  integrals over the $U_B(1)\times U_S(1)$ group with the weight function 
$\exp {(S[\phi_B,\phi_S,T,V_C])}$ 
which has only one canonical volume parameter $V_C$. 
Such an effective common volume parameter accounts simultaneously for the exact conservation of the baryon number and strangeness in a system and quantifies yields of hadrons carrying baryon and strangeness quantum numbers \cite{Braun-Munzinger:2003pwq}. 

We find,  an approximate  linearity of canonical correlation  volume as a function of the multiplicity as seen in Fig.~\ref{fig:radius}.
Thus, we fit the baryon canonical volume
as a function of charged particle multiplicity with a linear function,  
\begin{align}
    \begin{split}
    V_C &\simeq 27.3 + 2.9 \times \frac{dN_{ch}}{d\eta} 
\label{eq:volumec}
\end{split}
\end{align}
The above parametrization is also shown in Fig.~\ref{fig:radius} as dotted lines. 
{We note, however, that such a linear fit is valid only in the $dN_{ch}/d\eta$ range given by the above considered data in pp and pA collisions.} 
From the  fits to acceptance volume $V_A$ made for each multiplicity bin it was also shown  that it can be  well parameterised as linear function of  $dN_{ch}/d\eta$, as  \cite{Cleymans:2020fsc}:  
\begin{align}
    \begin{split}
    V_A &\simeq 1.55 + 3.0 \times \frac{dN_{ch}}{d\eta} .
\label{eq:volume}
\end{split}
\end{align}
 {This linear dependence of $V_A$ is only valid for $dN_{ch}/d\eta$ values greater than two.}


Having established the $dN_{ch}/d\eta$ dependence of volume parameters $V_A$ and $V_C$, and fixing the chemical freeze-out temperature at  
$ T = 156.5$
MeV,  the yields of 
(multi-) baryon states  
are  obtained  in the BCE from Eq.~\ref{eq:particle}.  
In the actual calculations the conservation of all other charges is included in the grand canonical ensemble with vanishing chemical potentials.

\begin{figure}[h!]
\centering
\includegraphics[scale=0.38]{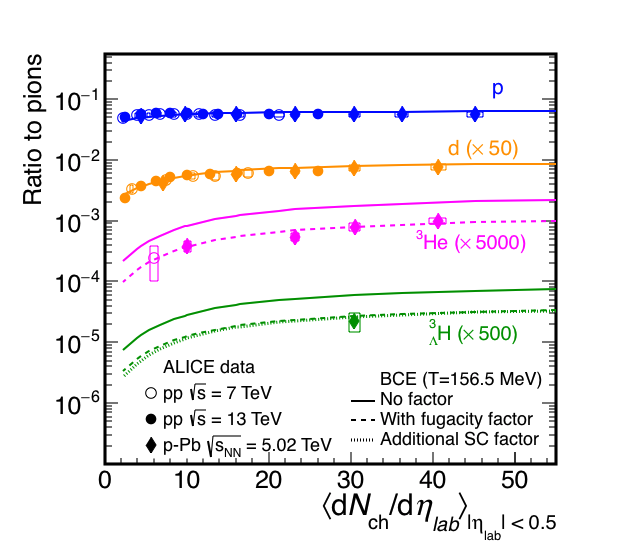}
\includegraphics[scale=0.38]{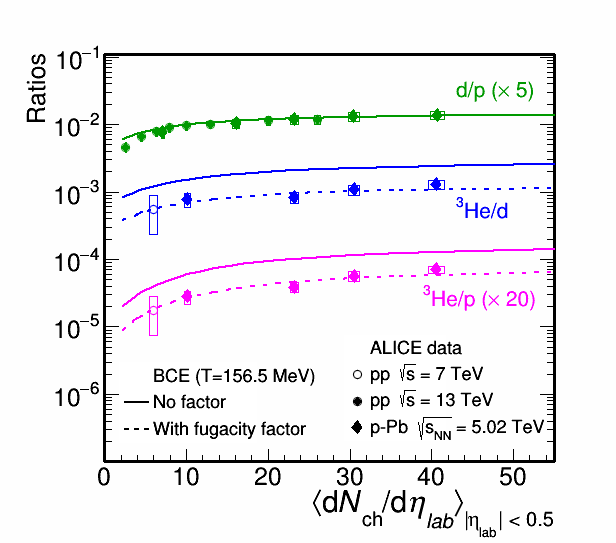}
\caption{Top: Ratios of protons,  deuterons, $^3$He and  $^3_\Lambda$H to  pions
  as a function of charged particle multiplicity.
  Bottom: The yield ratios d/p, $^3$He/d, and $^3$He/p as a function of charged particle multiplicity.
The various symbols represent the experimental data\cite{ALargeIonColliderExperiment:2021puh,ALICE:2020foi,ALICE:2019bnp,ALICE:2015wav,
ALICE:2017xrp,ALICE:2019fee} while the solid lines are the predictions from present  analysis
in  the BCE thermal model. The dashed lines (for ratios involving $^3$He and $^3_\Lambda$H) represent the same predictions but rescaled with a factor $\gamma=0.45$ obtained from the linear fit in Fig.~\ref{fig:yields}(c). The dotted line represents the effect of exact strangeness conservation in addition to baryon-canonical effects and fugacity factor for $^3_\Lambda$H.
}
\label{fig:ratios}
\end{figure}

The model results  as a function of charged particle multiplicity are shown in Figs.~\ref{fig:yields} and \ref{fig:ratios} and compared with data. The various symbols
represent the experimentally measured yields in pp and pA collisions from the ALICE
experiment~\cite{ALargeIonColliderExperiment:2021puh,ALICE:2020foi,ALICE:2019bnp,ALICE:2015wav,
ALICE:2017xrp,ALICE:2019fee,ALICE:2020nkc,ALICE:2016fzo}.
The quality  of  the BCE model  description of the data is quantified in Figs.~\ref{fig:yields}b and~\ref{fig:yields}c by showing the ratios of experimental data to model results.


The pions, protons and deuterons production   is  well
described by the BCE model for all multiplicities considered. This is very transparent  in Fig.~\ref{fig:yields} for  particle yields  and in Fig.~\ref{fig:ratios} for  their ratios.  In pA collisions the above yields data are consistent with the model results within  experimental uncertainties.  This is also the case for pp data  with   $dN_{ch}/d\eta>10$. For lower multiplicities in pp collisions the agreement is  at the level of  two standard deviations. 

The data for proton, deuteron and $^3$He capture also basic properties of the BCE model, namely the so called 
{\it canonical suppression effect} 
\cite{Braun-Munzinger:2003pwq,Hamieh:2000tk},
i.e. the suppression of  production yields of baryonic  states in low-multiplicity relative to high-multiplicity events. In  addition, this suppression increases with  the baryon number of the state. This property is very transparent in Fig.~\ref{fig:ratios} where the yields of protons, deuterons, and  $^3$He are normalized to pion yields, effectively suppressing the dependence on the volume parameter $V_A$.

Although, the qualitative trend of $^3$He production as a function of multiplicity is nicely predicted by the BCE
model, nevertheless the model results overpredict the 
observed yields. This is transparent in Fig.~\ref{fig:yields} for $^3$He yields as well as in their  ratios to pions,  protons and deuterons as shown in Fig.~\ref{fig:ratios}.   
It is particularly interesting to note,  that for $dN_{ch}/d\eta <50 $  the above  deviations of the BCE model results from  data are independent of  the charge particle multiplicity.  Indeed, in   Fig.~\ref{fig:yields}(c) we see that the ratio of experimental yields over the BCE  model predictions for $^3$He and even for $^3_\Lambda$H, is within uncertainties constant. 
This ratio is fitted to be, 
$\lambda=0.45 \pm 0.03$.
When rescaling  the model results  for $^3$He and $^3_\Lambda$H with this factor $\lambda$  (see dashed lines in Figs.~\ref{fig:yields} and \ref{fig:ratios})  the data are nicely reproduced 
for all values of $dN_{ch}/d\eta$.

We note, however, that since $^3_\Lambda$H carries a strange quantum number $|S|=1$, thus in small multiplicity events its yields are also subjects of additional suppression due to exact strangeness conservation (SC). The resulting strangeness suppression of $^3_\Lambda$H yields is also quantified in Figs.~\ref{fig:yields} and \ref{fig:ratios},  and is seen to be small in the parameter range considered. At the measured value of $^3_\Lambda$H  the strangeness  suppression is at the percentage level whereas at lower $dN_{ch}/dy $ it increases up to  $\approx 15 \%$. Such suppression is hardly visible on log plots.

The above-observed differences between the BCE model predictions and $^3$He,   $^3_\Lambda$H yields data by a constant multiplicative factor can be interpreted as being due to {\it deviations from chemical equilibrium}. In events with small $dN_{ch}/d\eta$
 the yields of multi-baryons states such as $^3$He and $^3_\Lambda$H appear in thermal but not in chemical equilibrium.  This can be quantified by the off-chemical equilibrium fugacity factor $\lambda$  which in the present case is nearly independent of charged particle multiplicity.  Considering, however, that in central Pb-Pb collisions the yields of  $^3$He and  $^3_\Lambda$H are found to be consistent with the HRG model results in chemical equilibrium, one expects that the above fugacity parameter must depend on $dN_{ch}/d\eta$ and will converge to unity for sufficiently large multiplicities.




\section{Conclusion}
We have analyzed the yields  of protons,
deuterons, $^3$He,  and  $^3_\Lambda$H  in the thermal fireball constrained by an exact baryon number conservation. 
We have applied the hadron resonance gas (HRG)  model formulated in the canonical ensemble concerning baryon number conservation, including contributions of all baryons and multi-baryon states. The model predictions have been  compared with  recent  yields data at mid-rapidity obtained by the ALICE Collaboration 
in  pp  and pPb collisions at $\sqrt{s}=13$ TeV and $\sqrt{s_{NN}}=5.02$ TeV, respectively. We have focused on charged particle multiplicity ($dN_{ch}/d\eta$) dependence of protons and light-nuclei production. The fireball thermal parameters, 
the freeze-out temperature, and the volume at mid-rapidity, as well as, its $dN_{ch}/dy$ dependence,  were taken the same as obtained in the HRG model analysis of strange particle production in the corresponding system \cite{Cleymans:2020fsc}.
Thus the only parameter left in this analysis was the correlation volume $V_C$ of an exact baryon number
conservation.
The extracted value of $V_C$ from the fit to pion,  proton {and deuteron} yields is nearly linearly-dependent on $dN_{ch}/d\eta$, and  for $2<dN_{ch}/d\eta<50$.  

We have shown that the observed yields of protons and deuterons and their $dN_{ch}/d\eta$ dependence are  well quantified   by model predictions. Also the qualitative trend  of data, i.e. the relative suppression of baryon yields with decreasing  $dN_{ch}/d\eta$ and its increase with the baryon content of the state is well reproduced by the thermal model with exact baryon-number conservation.  However, on the quantitative level,  the yields of $^3$He,  and  $^3_\Lambda$H  are  over-predicted by the  constant  multiplicative factor  which  has been  interpreted as the off-chemical equilibrium   fugacity factor. 
%
%
Thus, in small systems with $dN_{ch}/d\eta<50$ the  yields of 
hadronic states  carrying baryon number $|B|=1$ and 2 appear  to be consistent with  multiplicities expected in thermal and chemical equilibrium, where as light nuclei like  $^3$He,  and  $^3_\Lambda$H are lacking chemical equilibrium population. This is, however,  not the case for large systems like in central Pb-Pb collisions where   the yields of  nuclei and anti-nuclei including (anti-)hypernuclei  have been shown to  be  consistent with chemical equilibrium thermal model predictions. 

With the forthcoming data in these, including production of $^4$He and $^3_\Lambda$H  at  $dN_{ch}/dy<40 $,  the above conjecture  of possible off-chemical equilibrium effects in the light-nuclei production in small systems can be verified further or their thermal production in small multiplicity events can be excluded.

\section{Acknowledgments}
This work was started in collaboration with Jean Cleymans who passed away in February 2021. This  paper is dedicated to Jean. N.S. acknowledges the support of SERB Research Scientist research grant (D.O. No. SB/SRS/2020-21/48/PS)
 of the Department of Science and Technology, Government of India. N.S. acknowledge the discussions with B. Hippolyte.
  L.K. acknowledges the support received from NISER, Bhubaneswar, India and support from the research project No. SR/MF/PS-02/2021-PU (E-37120) of the Department of Science and Technology, Government of India.
  K.R.  acknowledges the support by the Polish National Science Center (NCN) under  Opus grant no.  2018/31/B/ST2/01663, and  
 partial support of the Polish Ministry of Science and Higher Education. K.R. also acknowledges   fruitful discussions with  P.  Braun-Munzinger, B. Friman, A. Rustamov  and J.  Stachel.

\newpage
\bibliographystyle{elsarticle-num}
\bibliography{references_cprs}
\end{document}